# Groove-rolling as an alternative process to fabricate Bi-2212 wires for practical applications

A Malagoli[1], V Braccini[1], M Vignolo[1], X Chaud[2] and M Putti[1,3]

[1]CNR-SPIN Corso Perrone 24, 16152 Genova, Italy
[2]CNRS LNCMI UJF-UPS-INSA, 25 rue des Martyrs BP.166 38042, Grenoble Cedex 9, France
[3]Physics Department, University of Genova, Via Dodecaneso 33, 16146, Italy

E-mail: andrea.malagoli@spin.cnr.it

**Abstract** $Bi_2Sr_2CaCu_2O_{8+x}$ (Bi-2212) superconducting long-length wires are mainly limited in obtaining high critical currents densities ($J_C$) by the internal gas pressure generated during the heat treatment, which expands the wire diameter and dedensifies the superconducting filaments. Several ways have been developed to increase the density of the superconducting filaments and therefore decreasing the bubble density: much higher critical currents have been reached always acting on the final as-drawn wires. We here try to pursue the same goal of having a denser wire by acting on the deformation technique, through a partial use of the groove-rolling at different wire processing stages. Such technique has a larger powders compaction power, is straightforwardly adaptable to long length samples, and allows the fabrication of samples with round, square or rectangular shape depending on the application requirements. In this paper we demonstrate the capability of this technique to increase the density in Bi-2212 wires which leads to a three-fold increase in $Jc$ with respect to drawn wires, making this approach very promising for fabricating Bi-2212 wires for high magnetic field magnets, i.e. above 25 T.

## 1. Introduction

Scientific applications involving magnet construction for the energy upgrade of Large Hadron Collider (LHC), for fusion (ITER), or for Nuclear Magnetic Resonance (NMR) magnets require superconducting conductors capable of generating - once wound as magnets - magnetic fields above 20 T [1,2]. Low temperature Superconductors (LTS) such as NbTi and $Nb_3Sn$ – which have been by now pushed up to their ideal performances – have intrinsic limitations in magnetic field: their critical current density $J_C$ has a strong dependence on magnetic field due to their relatively low upper critical field ($H_{c2}$) values – being 15 and 30 T, respectively – while High Temperature Superconductors (HTS) such as $YBa_2Cu_3O_{7-x}$ (YBCO), $(Bi,Pb)_2Sr_2Ca_2Cu_3O_{10-x}$ (Bi-2223) and $Bi_2Sr_2CaCu_2O_{8+x}$ (Bi-2212) show flat $J_C$ behaviour up to field above 30 T due to $H_{c2}$ exceeding 100 T [3].

For long time the bottleneck to the application of Bi-2212 wires for magnet construction has been the significantly lower $J_C$ of long lengths wires with respect to short samples. Previous reports [4, 5, 6, 7] have shown that the main reason for the degradation of critical current in long-length Bi-2212 wires are the voids or gas filled bubbles generated by agglomeration of porosity in the cold-worked wires during the heat treatment, which expand the wire diameter and dedensifie the superconducting filaments. Furthermore, the presence of impurities (e.g. C transforming to $CO_2$) can enhance the internal pressure increasing the void fraction and further deteriorate $J_C$. Different routes have been proposed to increase the density of the Bi-2212 filaments and therefore diminish the bubble density and / or size and as a consequence increase the critical current along the conductor. All of them act on the final as-drawn wire: $I_C$ was doubled through the use of a 2 GPa cold isostatic pressure (CIPping) [8] and through swaging [9] after drawing, while very high (up to 100 bar) over pressure (OP) applied during heat treatment was able to generate an eight-fold increase in the engineering critical current density $J_E$ [10] in wires with densities higher than 95%.

All the described approaches, though reaching extremely appealing $J_C$ values, might not be as simple and straightforward over long lengths and coils.

Aiming at developing a cheap and easily industrially scalable technique, we studied a different approach to densify the wire at an earlier stage using a different deformation technique in making Bi-2212 conductors [11]. We explored the groove-rolling process, whose peculiar characteristic compared to drawing is the difference in deformation forces exerted on the metallic sheath: during the groove rolling, the component of the force transversal with respect to the wire axis is higher resulting in a larger powders compaction capability. A similar approach has been successfully applied to *ex-situ* $MgB_2$ wires production [12, 13], where the density and thus the connectivity issue is one of the crucial points in reaching high critical currents as well. Moreover this process is very versatile: once optimized, going to the production of long wires to make coils is immediate. Furthermore, samples can be prepared in different shapes – not only round, but squared and rectangular as well – which can have some advantages. For example, in a solenoid type winding, having square or rectangular shaped conductor allows a better compaction of the turns drastically reducing the voids space between them; moreover with a rectangular wire it is possible to avoid unwanted conductor twisting during the winding process: in any case the geometry of such a wire has to be far away from being a tape, in order to maintain the peculiar Bi-2212 isotropy.

In this paper we present a series of wires with different architectures prepared by groove-rolling. Their transport properties are shown and compared not only with those of our as-drawn wires but also with the performances of the present commercial conductors. The characterization of a new rectangular wire is shown as well, together with a comparison between the critical current densities measured in open- and closed-ends samples, demonstrating a clear effect in terms of powder densification of this alternative deformation process.

## 2. Experimental

Several Bi-2212 wires differing in architecture and cold-working process were realized and analysed in this work (the list of the samples is reported in Table 1). The Powder-In-Tube method was used for all samples starting from filling an Ag tube with Nexans granulate powder whose overall composition was $Bi_{2.16}Sr_{1.93}Ca_{0.89}Cu_{2.02}O_X$. The tube outer (OD) and inner (ID) diameters were 15 and 11 mm respectively. After drawing, the obtained monofilamentary wire was used as basic element for all wires. The preparation of the samples called SPIN_D and SPIN_G has already been described in our previous work [7]: the hexagonal shaped monofilamentary wire was cut in 55 pieces and restacked in a further 15/11 (OD/ID) mm Ag tube.

At this point the deformation process was differed for the two wires. For the sample SPIN_D the restacked tube was drawn, hexagonal shaped and cut in 7 pieces in order to be restacked again in a 10/7.5 mm (OD/ID) Ag/Mg alloy tube. After a further drawing process, a 55 x 7 filaments round wire with a diameter of 0.8 mm was obtained. For the SPIN_G sample the restacked tube with 55 filaments was worked with a groove-rolling machine.

To prepare the three samples SPIN_1, SPIN_2 and SPIN_3, 37 monofilaments were restacked in a 13/11 (OD/ID) mm Ag tube. After few drawing steps, such second tube was groove-rolled and hexagonal shaped: 18 pieces of it were restacked in a further Ag tube (9.5/7.5 mm) together with one central Ag rod. Starting from this last composite tube, the three samples were realized changing the deformation process: sample SPIN_1 was prepared only with the drawing process to obtain a round wire of 1.1 mm in diameter; for the sample SPIN_2 the tube was drawn down to a diameter of 2.2 mm and then groove-rolled to a square-shaped wire with a final size of 1.1 x 1.1 mm$^2$; finally for the sample SPIN_3 the tube was drawn down to a diameter of 4.2 mm and then groove rolled to a square-shaped wire of 1.1 x 1.1 mm$^2$.

For the last sample, SPIN_R, 85 monofilamentary elements were restacked in a 13/11 (OD/ID) mm Ag tube which was groove-rolled, hexagonal shaped and cut in 7 pieces to be restacked again in a 9.5/7.5 mm Ag tube. Such composite tube was groove-rolled again to obtain a square wire of 1.2 x 1.2 mm$^2$. From that, by means a flat-rolling machine, a 85 x 7 filaments rectangular wire of 1.5 x 1 mm$^2$ in size was realized. Table 1 summarizes each sample preparation steps. Figure 1 shows images of transverse cross sections of all samples, carefully polished with SiC paper and, as a last step, with a diamond paste. The cross-sections were imaged in an optical microscope which gave high contrast digital images. These images were analysed with ImageJ so as to extract the area fraction of the Ag matrix and filament packs for evaluation of the effective area of the filaments.

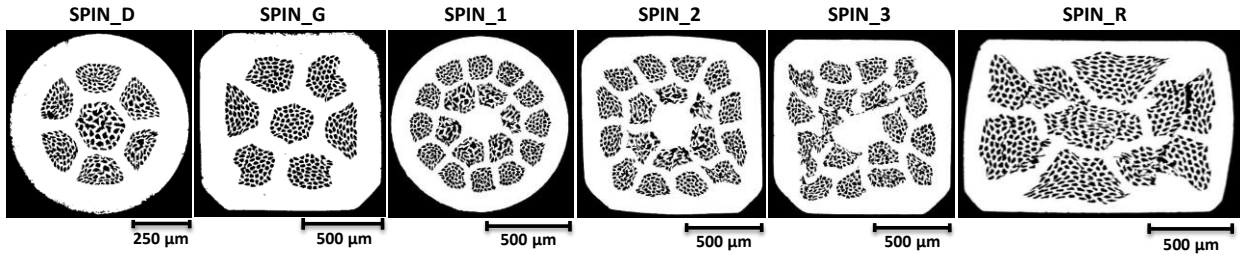

**Figure1.** Images of transverse cross sections of all samples produced by SPIN

12 cm long wires cut from each sample were heat treated in 1 bar flowing O$_2$ in a tubular furnace with a homogeneity zone (± 0.5 °C) of 30 cm using the standard heat treatment schedule[14]. Actually no temperature optimization was performed on our tubular furnace, especially regarding the so called $T_{MAX}$: in this work we adopted $T_{MAX}$ = 888 °C.

Regarding the SPIN_R sample, a piece of it was left with the ends open while in another piece the ends were sealed by dipping them into liquid silver before the heat treatment. The time of this dip was a compromise between being long enough to obtain a good seal, and short enough to avoid melting of the external sheath of the wire. After the heat treatment no leakages were observed at the edges of the sealed SPIN_R sample.

Transport critical currents were measured on 10 cm long samples by means of two four-probe systems: the first using a 7 T split-coil magnet at 4.2 K installed at SPIN and the second using a 13 T magnet at 4.2 K at High Magnetic Field Laboratory in Grenoble (FR). In both cases the field was applied perpendicular to the wire axis. The voltage taps were located 1-1.2 cm apart in the centre of the 10 cm long samples. A criterion of 1 µV/cm was used. The critical current density $J_C$ was calculated taking into account the oxide filaments area measured by image analysis on the unreacted wire as it is used in literature [15].

**Table 1**. Sample preparation steps

| wire | 1st restack | 2nd restack | final shape & size |
|---|---|---|---|
| SPIN_D | 55 fil drawn | 55 x 7 fil drawn | Round ⌀ 0.8 mm |
| SPIN_G | 55 fil groove-rolled | 55 x 7 fil groove-rolled | Square 1.1x1.1 mm$^2$ |
| SPIN_1 | 37 fil groove-rolled | 37 x 18 fil drawn | Round ⌀ 1.1 mm |
| SPIN_2 | 37 fil groove-rolled | 37 x 18 fil drawn to 2.2 mm / groove-rolled | Square 1.1x1.1 mm$^2$ |
| SPIN_3 | 37 fil groove-rolled | 37 x 18 fil drawn to 4.2 mm / groove-rolled | Square 1.1x1.1 mm$^2$ |
| SPIN_R | 85 fil groove-rolled | 85 x 7 fil groove / flat-rolled | Rectangular 1.5x1.0 mm$^2$ |

## 3. Results

In Figure 2 the transport critical current densities $J_C$ at 4.2 K in applied field up to 13 T are shown for the samples SPIN_1, SPIN_2 and SPIN_3 compared to those of the samples SPIN_D and SPIN_G measured up to 7 T. A positive trend in $J_C$ is evident passing from SPIN_1 to SPIN_2 and, more remarkably, to sample SPIN_3, where even a three-fold improvement in $J_C$ can be observed. It is therefore clear as well how such $J_C$ behaviour is directly correlated with the changes in the mechanical deformation process. All the samples have the same starting architecture made by the same 37-filaments groove-rolled bundles. The differences lie in the deformation of the second restack, where the groove-rolling process was introduced just at the last steps in SPIN_2 and for almost half of the working process in SPIN_3. These results are compared with those obtained in sample SPIN_D and SPIN_G described in [11]. The $J_C$ values measured in SPIN_1 are the same as in SPIN_D: it seems that having applied the groove-rolling at the first restack in SPIN_1 was made useless by applying the drawing for the rest of the deformation. On the other side the fact that also SPIN_G and SPIN_3 have the same $J_C$ values, means that the groove-rolling steps performed after the drawing were enough in SPIN_3 to get the high $J_C$ measured in the groove-rolled SPIN_G.

In a second experiment we measured the performances of the rectangular 85 x 7 filaments SPIN_R wire. As described in the introduction of this paper, we think that the rectangular cross section shape could meet, better than round shape, the magnet manufacturer requirements for several applications. To simulate the behaviour of a long coil length we sealed the ends of a sample cut from such wire and performed a comparison between the transport properties of this samples and another with open ends. As described in [7] sealing the sample ends is a way to make it quite difficult for the residual inner gas, provoking the bubbles formation, to escape during the heat treatment, as it occurs in a very long wire. Such comparison is useful in understanding how the gas amount inside the wire, and thus the Bi-2212 powder density, can influence the transport performances of a winding length conductor.

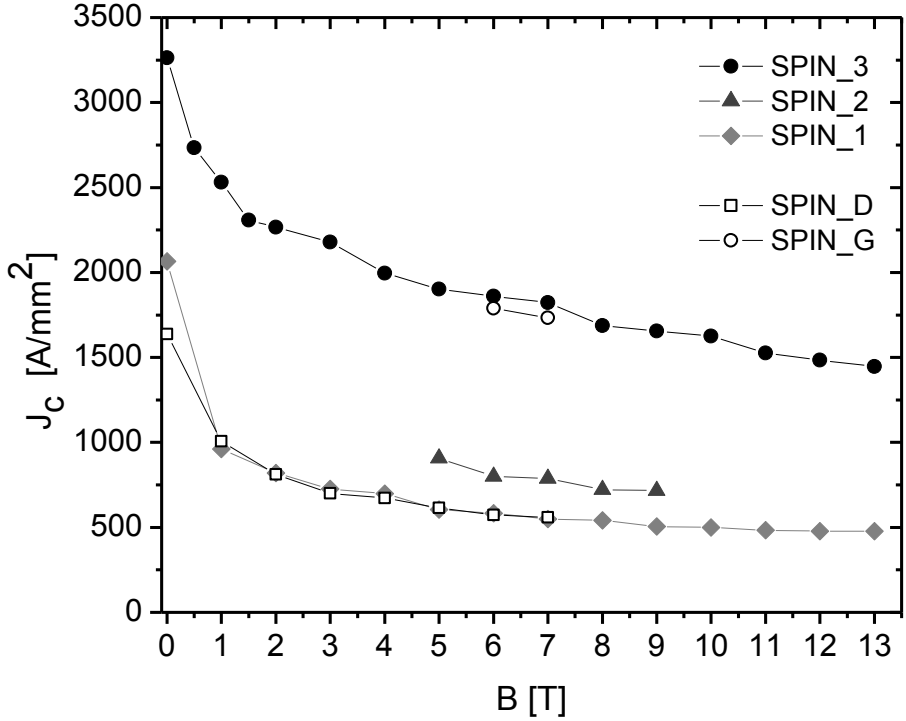

**Figure 2.** Comparison between the transport critical current densities $J_C$ vs. $B$ showing the effect of the groove-rolling steps on the sample performances

In Figure 3 the transport $J_C$ measured up to 7 T for both samples SPIN_R-O (open ends) and SPIN_R-C (closed ends) is reported. For the SPIN_R-C sample a $J_C$ reduction of 28% and constant in field is observed. The resulting sealed wire is shown in the inset.

## 4. Discussion

We showed how it is possible to obtain a three-fold increase in $J_C$ through the groove-rolling process during the deformation: our 1.1 mm$^2$ square 37 x 18 groove-rolled wire reaches a $J_C$ higher than 1900 A/mm$^2$ at 5 T (see Figure 2).

The principal aim of this work was to better analyse the effect of this working process. Groove-rolling is an already industrialized technique nowadays employed, for example, in the production of commercial magnesium diboride superconducting wires and it is worth to explore its potential in enhancing the Bi-2212 density and thus its critical current density. Observing the results shown in Figure 2 a positive trend in terms of $J_C$ is clearly given by progressive addition of groove-rolling steps. On the other side, the comparison between sample SPIN_1 and SPIN_D, which have the same $J_C$, shows how the drawing is not only less efficient in powder densification then groove-rolling, but also it provokes a dedensification of the filament bundles previously worked by groove-rolling, bringing back the $J_C$ of SPIN_1 to the SPIN_D values. It is also important to remind how in our previous paper [11] it was shown that, compared to drawing, groove rolling makes the final wire quality less sensitive to inhomogeneities of initial precursor as demonstrated for example by the absence of hard particles inside the cold-worked wire filaments. However, we are still far away from having a proper optimization of such process, especially regarding the relationship between the deformation steps and the wire architecture and shape, filaments size, filling factor and sheath thickness.

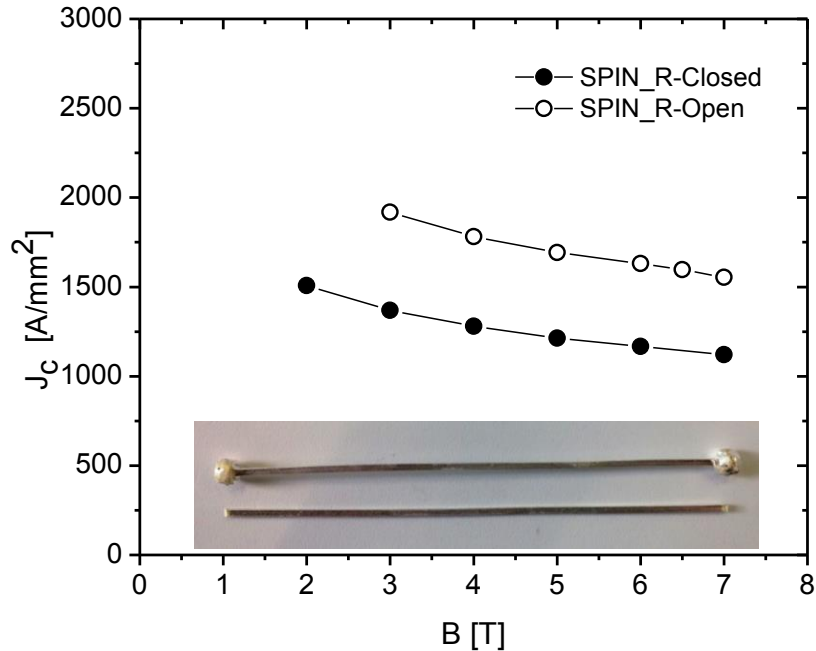

**Figure 3.** Transport $J_C$ measured up to 7 T for both samples SPIN_R-O (open ends) and SPIN_R-C (closed ends). For the SPIN_R-C sample a $J_C$ reduction of 28% and constant in field is observed. In the inset the resulting sealed wire is shown after the heat treatment

Nevertheless, it is worth doing a comparison in terms of $J_C$, architecture, and filling factor with the industrially produced conductors. Wires with architectures similar to ours – but always round – were produced by Nexans Korea. In particular, round wires with 385 (55x7) and 595 (85x7) filaments were prepared [16]. Such samples show $J_C$ between 2100 and 1700 A/mm² at 0 T and 4.2 K: if we compare such values with those reported in Figure 2, we see that our drawn samples lie in the very same range. Wires produced by Supercon. Inc. have a single restack and a very high filling factor: a very deep work has been performed on the heat treatment on such wires very recently, and the $J_C$ has been enhanced from around 700 to a maximum of 1600 A/mm² at 5 T through the saw-tooth processing [17]. The best Bi-2212 wires available so far are those fabricated by Oxford Superconducting Technology [18]: as already mentioned in the introduction, the properties of such wires have been extensively studied and a lot has been done to increase their density through CIPping, swaging and heat-treatment with an over pressure up to 100 bar, reaching 2500 A/mm² at 20 T and almost 4000 A/mm² at 5 T [10], well above the requirements for high field magnets. It is known that the overall cation composition of precursors strongly affects the final performances [19], therefore it is important to note that the overall composition of granulated precursors from Nexans used in this work is the same to that used by all of these wire manufacturers. In Figure 4 we report a comparison between the behaviour of $J_C$ with the magnetic field for the groove-rolled sample SPIN_3 and the as-drawn sample OST – which we have heat treated at SPIN - with the same number of filaments 37 x 18: as also reported in Table 2, there is agreement in the measured values indicating a proper heat treatment performed. In Figure 4 we also show the optical cross sections of the two wires. First of all, we notice that our sample has a much lower fill factor – it is around 17% for all our wires – while the OST sample has 28% of superconducting fraction, reaching 32% in the 85 x 7 configuration [9]. Furthermore, the cross-section is not regular: some of the bundles look damaged by the deformation process, and might be interrupted over their length. We therefore need to optimize the Ag tubes thickness – in particular thinner tubes could be more successfully employed – and also the final filaments size. Furthermore, at first we replicated the architecture which was optimized for the fabrication of round wires through drawing: we need to find now a new configuration which is more suitable for square samples, i.e. starting from square instead of round tubes.

In our first report of Bi-2212 wires prepared through groove-rolling, we evaluated the superconductor mass density, and found that the groove-rolled wire had an increase of about 30% over the Bi-2212 of the drawn sample [11]. We here support the idea that a stronger mechanical deformation increases the density of the superconducting phase through a different experiment, i.e. heat treating the same wire with ends either open or closed, following the idea that a closed-ends wire can 'mimic' what happens in long-length samples [4]. We have reported in Figure 3 the reduction of $J_C$ in the rectangular-shaped sample when heat treated with closed-ends, which is of about 28% over the whole measured magnetic field range. In Table 2 we show for comparison data from other reports in which this same experiment has been performed: the critical current is reduced by more than three times on the as-drawn OST wire. This means that in our case we still have some residual gas - and therefore we still have margin for improvement - but not as much as in drawn wires.

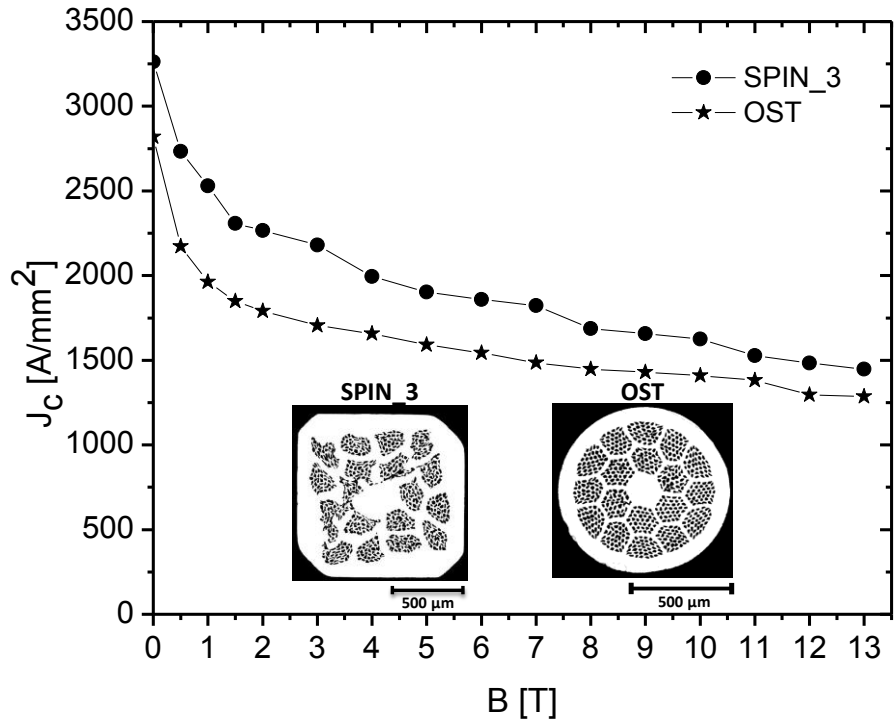

**Figure 4.** Comparison between the $J_C$ of our SPIN_3 sample and the as-drawn OST sample. In the inset, showing both cross section, it is possible to observe the differences regarding the filaments uniformity and fill factor.

**Table 2.** Specifications and critical current densities at 4.2 K and 5 T of SPIN and commercial as-drawn wires

| wire | architecture | Size [mm or mm$^2$] | $J_C$ [A/mm$^2$] open-ends | $J_C$ [A/mm$^2$] closed-ends |
| --- | --- | --- | --- | --- |
| OST | 37 x 18 | ⌀ 0.8 | 1667 [8] | 500 [10] |
| OST | 85 x 7 | ⌀ 1.0 | 1520 [9] | |
| OST $_{HeatTreated@SPIN}$ | 37 x 18 | ⌀ 0.8 | 1590 | |
| SPIN_R | 85 x 7 | 1 x 1.5 | 1693 | 1214 |
| SPIN_3 | 37 x 18 | 1.1 x 1.1 | 1903 | |

Besides that, the importance of the optimization of a wire with a rectangular cross section lies in some technical advantages. Especially for solenoid type winding, as NMR magnets, having a rectangular shaped conductor allows a better compaction of the turns drastically reducing the voids space between them and avoids unwanted conductor twisting during the winding process. Furthermore, looking at the cross section of SPIN_R in Figure 1 it is possible to observe that groove-rolling makes the corners smooth and this is a very desirable condition to avoid damages in the insulation. Finally, the geometric ratio 1.5/1 makes this wire very far from being a tape, maintaining the peculiar Bi-2212 isotropy.

The $J_C$ values are not high enough yet to fully satisfy the high field magnets requirements, especially if compared with those of the OP samples, but already comparable with those of well optimized commercial as-drawn wires. As it is well known getting uniform filaments with a proper size and a higher fill factor is necessary to develop wires with high $J_C$ and, even more important concerning practical applications, high engineering critical current density $J_E$: our SPIN_3 and SPIN_R wires, for instance, have a $J_E$ of 360 and 290 A/mm$^2$ respectively at 4.2 K and 5 T, a factor about 2 lower than that requested. The OST samples have an average single filament diameter of 15 μm which is the optimum size for the best $J_C$ [20] while SPIN_R, for example, has still an average filament diameter of about 25 μm. Finally also the thickness of the internal sheaths plays a crucial role in particular in enhancing the fill factor: the internal Ag sheath thickness in all OST samples is about 10 μm while in our samples, for example again in SPIN_R, such thickness is about 20 μm.

The next step will be to start using groove-rolling process since the fabrication of the mono-core wire and for the whole wire deformation in order to definitely avoid any drawing step. The challenge to be faced will be to obtain a very regular, uniform and dense conductor with suitable transport properties for some industrial high magnetic field applications.

## 5. Conclusions

We prepared Bi-2212/Ag wires with different architectures and shapes – i.e. square and rectangular - through groove-rolling. We found that it is possible to obtain a three-fold increase in $J_C$ with respect to the drawing through this alternative deformation process: our 1.1 mm$^2$ square 37 x 18 groove-rolled wire reaches a $J_C$ higher than 1900 A/mm$^2$ at 5 T. This result is remarkable because we still have a strong margin for improvement, given that the process is not optimized yet in terms of geometry of the restacks, thickness of the external and internal Ag tubes, filament average dimensions, and in general filling factor. Nevertheless, the increase in the critical current is entirely due to the deformation process which is able to densify the superconducting powders and therefore diminish the bubble density and / or size, while all the other methods which have been proposed and exploited to overcome this issue act on the final as-drawn wire. The clear effect in terms of powder densification of this alternative deformation has been shown through the comparison between the critical current densities measured in the open- and closed-ends rectangular wire: the fact that the reduction we register is remarkably less than what observed in drawn commercial wires is a strong indication of a higher density.

We believe that the use of groove-rolling instead of drawing to fabricate Bi-2212 wires is an appealing way to obtain denser and long-length wires with $J_C$ values appealing for some high field applications.

**Acknowledgements**
High field measurements at LCMI-Grenoble have been supported by the EU contract number 228043